\documentclass[tighten,times,twocolumn]{aastex61}

\shorttitle{Revival of the magnetar PSR~J1622$-$4950}
\shortauthors{Camilo et~al.}

\usepackage{graphicx}
\usepackage{epsfig}
\usepackage{amsmath,amssymb,amsthm}
\usepackage{lineno}
\usepackage{natbib}

\newcommand{\src}{PSR~J1622$-$4950}
\newcommand{\parkes}{{Parkes}}
\newcommand{\meerkat}{{MeerKAT}}
\newcommand{\xmm}{\textit{XMM-Newton}}
\newcommand{\swift}{\textit{Swift}}
\newcommand{\chandra}{\textit{Chandra}}
\newcommand{\nustar}{\textit{NuSTAR}}

\begin{document}

\title{Revival of the magnetar \src: observations with \meerkat,
\parkes, \xmm, \swift, \chandra, and \nustar\ }

\AuthorCallLimit=208

\author{F.~Camilo}
\affiliation{SKA South Africa, Pinelands 7405, South Africa}
\correspondingauthor{F.~Camilo}
\email{fernando@ska.ac.za}
\author{P.~Scholz}
\affiliation{National Research Council of Canada, Herzberg Astronomy
and Astrophysics, Dominion Radio Astrophysical Observatory, Penticton,
BC V2A 6J9, Canada}
\author{M.~Serylak}
\affiliation{SKA South Africa, Pinelands 7405, South Africa}
\affiliation{Department of Physics \& Astronomy, University of the
Western Cape, Bellville 7535, South Africa}
\affiliation{Station de Radioastronomie de Nan\c{c}ay, Observatoire
de Paris, CNRS, PSL Research University, Universit\'e Orl\'{e}ans,
F-18330, Nan\c{c}ay, France}
\author{S.~Buchner}
\affiliation{SKA South Africa, Pinelands 7405, South Africa}
\author{M.~Merryfield}
\affiliation{National Research Council of Canada, Herzberg Astronomy
and Astrophysics, Dominion Radio Astrophysical Observatory, Penticton,
BC V2A 6J9, Canada}
\affiliation{Department of Physics and Astronomy, University of
Victoria, Victoria, BC V8W 2Y2, Canada}
\author{V.~M.~Kaspi}
\affiliation{Department of Physics and McGill Space Institute,
McGill University, Montreal, QC H3A 2T8, Canada}
\author{R.~F.~Archibald}
\affiliation{Department of Physics and McGill Space Institute,
McGill University, Montreal, QC H3A 2T8, Canada}
\affiliation{Department of Astronomy and Astrophysics, University
of Toronto, Toronto, ON M5S 3H4, Canada}
\author{M.~Bailes}
\affiliation{Centre for Astrophysics and Supercomputing, Swinburne
University of Technology, Hawthorn, VIC 3122, Australia}
\affiliation{ARC Centre of Excellence for Gravitational Wave Discovery
--- OzGrav, Swinburne University of Technology, Hawthorn, VIC 3122,
Australia}
\author{A.~Jameson}
\affiliation{Centre for Astrophysics and Supercomputing, Swinburne
University of Technology, Hawthorn, VIC 3122, Australia}
\author{W.~van~Straten}
\affiliation{Centre for Astrophysics and Supercomputing, Swinburne
University of Technology, Hawthorn, VIC 3122, Australia}
\affiliation{Institute for Radio Astronomy \& Space Research,
Auckland University of Technology, Auckland 1142, New Zealand}
% ...

\received{17 October 2017}
\accepted{26 February 2018}
\submitjournal{{\em The Astrophysical Journal}}

\begin{abstract}
New radio (MeerKAT and Parkes) and X-ray (\xmm, \swift, \chandra,
and \nustar) observations of \src\ indicate that the magnetar, in
a quiescent state since at least early 2015, reactivated between
2017 March 19 and April 5.  The radio flux density, while variable,
is approximately $100\times$ larger than during its dormant state.
The X-ray flux one month after reactivation was at least $800\times$
larger than during quiescence, and has been decaying exponentially
on a $111\pm19$ day timescale. This high-flux state, together with
a radio-derived rotational ephemeris, enabled for the first time
the detection of X-ray pulsations for this magnetar. At 5\%, the
0.3--6\,keV pulsed fraction is comparable to the smallest observed
for magnetars. The overall pulsar geometry inferred from polarized
radio emission appears to be broadly consistent with that determined
6--8 years earlier.  However, rotating vector model fits suggest
that we are now seeing radio emission from a different location in
the magnetosphere than previously. This indicates a novel way in
which radio emission from magnetars can differ from that of ordinary
pulsars.  The torque on the neutron star is varying rapidly and
unsteadily, as is common for magnetars following outburst, having
changed by a factor of 7 within six months of reactivation.

\end{abstract}


\section{Introduction} \label{sec:intro}

Magnetars are the most magnetic objects known in the universe: a
class of neutron stars with high-energy emission powered by the
decay of ultra-strong magnetic fields, rather than through rotation
\citep{td95}.  This is particularly manifest during large X-ray
outbursts, when their luminosity exceeds that available from
rotational spin down.  The remarkable magnetar phenomenon provides
a unique window into the behavior of matter at extreme energy
densities.  Although all confirmed magnetars spin slowly ($\approx
1$--10\,s), it has been suggested that soon after birth, while
spinning at millisecond periods, they could be responsible for some
gamma ray bursts \citep[e.g.,][]{bgm17} or possibly fast radio
bursts \citep[e.g.,][]{mbm17}.

Observations of both magnetars and rotation-powered pulsars are
also uncovering links between the two populations.  \citet{crh+06}
discovered that magnetars can emit radio pulsations.  Magnetar radio
properties are often distinct from those of ordinary rotation-powered
pulsars, e.g., in having flat radio spectra \citep[e.g.,][]{crp+07}.
On the other hand, while radio magnetars have highly variable pulse
profiles, rotating vector model \citep[RVM;][]{rc69a} fits to
polarimetric observations often yield unexpectedly good results
that suggest a magnetic field geometry at the location of emission
not unlike that of ordinary pulsars (e.g., \citealt{crj+08}, but
see also \citealt{ksj+07}).  Conversely, distinct magnetar-like
outbursts, including short X-ray bursts and long-duration X-ray
flux enhancements, have now been observed from two pulsars formerly
classified as entirely rotation powered \citep[][]{ggg+08,akts16}.

Thus, more neutron stars that occasionally display magnetar behavior
surely lurk amidst the $\approx 2500$ known ``ordinary'' pulsars.
All the while, the careful study of the rotational and radiative
behavior of the two dozen confirmed magnetars \citep{ok14}\footnote{Catalog
at \url{http://www.physics.mcgill.ca/~pulsar/magnetar/main.html}},
coupled with remarkable theoretical progress, continues to advance
our understanding of these exceptional objects \citep[for a recent
review, see][]{kb17}.

Only four magnetars are known to emit radio pulsations. The first
to be identified, XTE~J1810$-$197, remained an active radio source
for approximately five years following the X-ray outburst that
resulted in its discovery, and has been radio-dormant since, for
nine years, while X-ray activity continues at a relatively low level
\citep{crh+16}.  Two others, 1E~1547.0$-$5408 and SGR~1745$-$2900,
have remained very active radio and high-energy sources \citep[e.g.,][and
references therein]{laks15,crt+17,mah17}.

\src, with rotation period $P=4.3$\,s, remains the only magnetar
discovered at radio wavelengths without prior knowledge of an X-ray
counterpart \citep{lbb+10}. At the time of discovery, its X-ray
flux was decaying exponentially from a presumed outburst in 2007,
and no X-ray pulsations could be detected \citep{ags+12}. Detectable
radio emission ceased in 2014 and, despite frequent monitoring, the
pulsar remained undetectable through late 2016 \citep{scs+17}.

Here we report on new multi-wavelength observations of \src, showing
that the magnetar is once again in a highly active state.  Observations
with the CSIRO Parkes telescope first detected resumed radio emission
on 2017 April 5.  Subsequent observations with the new Square
Kilometre Array South Africa (SKA SA) MeerKAT radio telescope have
been tracking the unsteady spin-down torque, and, as we describe
here, have enabled the first detection of X-ray pulsations for this
neutron star through the folding of X-ray photons collected with
\chandra\ and \nustar.  Further X-ray observations with \xmm\ and
\swift\ provide a fuller view of the spectral evolution of the star
following this most recent outburst.

\section{Observations} \label{sec:obs}

\subsection{Parkes} \label{sec:pksobs}

\subsubsection{Monitoring} \label{sec:mon}

Following the last observation reported in \cite{scs+17}, on 2016
September 16, we continued monitoring \src\ at the Parkes 64\,m
radio telescope. As before, the observations were performed with
the PDFB4 digital filterbank in search mode, typically at a central
frequency of 3.1\,GHz (recording a bandwidth of 1\,GHz), for $\approx
15$ minutes per session.

On 2017 April 26 we noticed during real-time monitoring of the
observations that single pulses were being detected from the pulsar.
Parkes underwent a planned month-long shutdown in May, and we started
monitoring \src\ with MeerKAT in late April (Section~\ref{sec:mkpsrobs}).

\subsubsection{Polarimetry} \label{sec:pol}

In order to compare the geometry of \src\ following its reactivation
in 2017 with that before its disappearance by early 2015, we have
made two calibrated polarimetric observations at the Parkes radio
telescope, using PDFB4 in fold mode.  These 40 minute observations
were performed at 1.4\,GHz (with 256\,MHz bandwidth, on 2017 August
5) and 3.1\,GHz (recording 1024\,MHz of bandwidth, on August 16),
using the center beam of the 20\,cm multibeam receiver and the
10\,cm band of the 1050cm receiver, respectively.

\subsection{MeerKAT} \label{sec:mkobs}

MeerKAT\footnote{\url{http://www.ska.ac.za/gallery/meerkat/}}, a
precursor to the Square Kilometer Array (SKA), is a radio interferometer
being built by SKA South Africa in the Karoo region of the Northern
Cape province, at approximate coordinates $21^\circ26'$ east,
$30^\circ42'$ south. The full array, scheduled to start science
operations in 2018, will consist of 64 13.5\,m diameter antennas
located on baselines of up to 8\,km.  The observations reported in
this paper were obtained during the commissioning phase and used a
subset of the array and some interim subsystems. As MeerKAT is a
new instrument not yet described in the literature, we provide here
an overview of the system relevant to our results.

\subsubsection{Receptors}

The MeerKAT dishes are of a highly efficient unshaped ``feed down''
offset Gregorian design. This allows the positioning of four receiver
systems on an indexing turret near the subreflector without
compromising the clean optical path. A receptor consists of the
primary reflector and subreflector, the feed horns, cryogenically
cooled receivers and digitizers mounted on the feed indexer, as
well as associated support structures and drive systems, all mounted
on a pedestal.  The observations reported here were performed at L
band. Averaged across this 900--1670\,MHz band, the measured system
equivalent flux density (SEFD) of one receptor on cold sky is
$\approx 460$\,Jy. UHF (580--1015\,MHz) and S-band (1.75--3.5\,GHz)
receiver systems are also being built (the latter by MPIfR).

The RF signal from the L-band receiver is transferred via coaxial
cables to the shielded digitizer package $\sim 1$\,m away.  The
digitizer samples the signal directly in the second Nyquist zone
without heterodyne conversion.  After RF conditioning and
analog-to-digital conversion, the 10-bit voltage stream from each
of two (horizontal and vertical) polarizations is framed into
$4\times10$\,Gbps Ethernet streams. These are concatenated onto a
single 40\,Gbps stream for transmission via buried optical fiber
cables to the central Karoo Array Processor Building (KAPB) located
$< 12$\,km away.  The one-pulse-per-second signal that allows precise
time stamping of voltage sample data and the sample clock frequencies
for the analog-to-digital converters (1712\,MHz for L band) originate
in the Time and Frequency Reference (TFR) subsystem located in the
KAPB. Ultimately, two hydrogen maser clocks and time-transfer GPS
receivers will allow time stamping to be traceable to within 5\,ns
of UTC. The observations presented here made use of an interim TFR
system with GPS-disciplined rubidium clocks, which provides time
stamps accurate to within 1\,$\mu$s of UTC.

\subsubsection{Correlator/Beamformer}

The MeerKAT correlator/beamformer (CBF) implements an FX/B-style
real-time signal processor in the KAPB. The antenna voltage streams
are coarsely aligned as necessary to compensate for geometric and
instrumental delays, split into frequency channels, and then phase
aligned per frequency channel, prior to cross-correlation (X) and/or
beamforming (B). This is done in the ``F-engine'' processing nodes,
where a polyphase filterbank is used to achieve the required
channelization with sufficient channel-to-channel isolation.

The CBF subsystem is based on the CASPER technology, which uses
commodity network devices (in the case of MeerKAT, 40\,Gbps Mellanox
SX1710 36-port Ethernet switches arranged in a two-layer CLOS network
yielding 384 ports) to handle digital data transfer and re-ordering
between processing nodes. The switches allow a multicast of data
to enable parallel processing. The processing nodes for the full
MeerKAT array will consist of so-called SKARAB boards populated
with Virtex~7 VX690T field-programmable gate arrays (FPGAs). The
interim CBF, used for the observations presented here, uses the
ROACH2 architecture populated with Virtex~6 SX475T FPGAs, and can
handle a maximum of 32 inputs, such as two polarizations for each
of 16 antennas.

For pulsar and fast transient applications, a tied-array beam is
formed in the B engines, which perform coherent summation on
previously delayed, channelized, and re-ordered voltage data. The
F and B engines handle both polarizations, but polarization calibration
still has to be implemented for tied-array mode, and the observations
presented here are based on uncalibrated total-intensity time series.

\subsubsection{Pulsar timing backend}

For all observations presented here, data from a dual polarization
tied-array beam split into 4096 frequency channels spanning 856\,MHz
of band centered at 1284\,MHz were sent to the pulsar timing backend.
This instrument is being developed by the Swinburne University of
Technology pulsar group. The hardware consists of two eight-core
servers, each equipped with four NVIDIA Titan~X (Maxwell) GPUs,
128\,GB of memory, a large storage disk, and dual-port 40\,Gbps
Ethernet interfaces, through which the beamformed data are received.
This allows the simultaneous processing of up to four tied-array
beams, although at present only one is provided by the B engines.

The beamformed voltage data stream is handled by a dedicated real-time
pipeline. First, the UDP packets are received and allocated to a
PSRDADA\footnote{\url{http://psrdada.sourceforge.net/}} ring buffer.
Next, the data are asynchronously transferred to the GPUs, dedispersed,
detected, and folded into 1024 phase bins by DSPSR \citep{vb11}
using a TEMPO2 \citep{hem06} phase predictor derived from the
pulsar's ephemeris.  Every 10\,s, a folded sub-integration is
unloaded from the GPUs to local storage in PSRFITS format \citep{hvm04}
for subsequent offline analysis.  Data acquisition and processing,
as well as control and monitoring of the backend, are handled by
SPIP\footnote{\url{http://github.com/ajameson/spip/}}, a C++, PHP
and python software framework that combines the individual software
tools mentioned above into a complete pulsar timing instrument. In
the observations presented here, the data were dedispersed incoherently,
although provision exists for coherent dedispersion.

\subsubsection{MeerKAT observations} \label{sec:mkpsrobs}

\begin{deluxetable*}{cccccc}[!t]
\tablecaption{X-Ray Observations of \src\ Since Its 2017 Outburst \label{tab:xrayobs}}
\tablecolumns{6}
\tablewidth{0pt}
\tabletypesize{\footnotesize}
\tablehead{
	\colhead{Telescope} & Energy & \colhead{ObsID} & \colhead{Date} & \colhead{Start Time} & \colhead{Exposure Time} \\ 
	\colhead{} & \colhead{(keV)} & \colhead{} & \colhead{(YYYY-Month-DD)} & \colhead{(MJD)} & \colhead{(ks)} 
}
\startdata
\swift\   & 0.2--10 & 00010071001 & 2017 Apr 27 & 57870.7 &   2.5 \\
\swift\   & 0.2--10 & 00010071002 & 2017 May 01 & 57874.7 &   5.0 \\
\swift\   & 0.2--10 & 00010071003 & 2017 May 05 & 57878.7 &   1.7 \\
\nustar\  & 3--79   & 80202051002 & 2017 May 07 & 57880.9 &  52.6 \\
\chandra\ & 0.3--10 & 19214       & 2017 May 08 & 57881.8 &  10.1 \\
\chandra\ & 0.3--10 & 19215       & 2017 May 23 & 57896.2 &  15.0 \\
\nustar\  & 3--79   & 80202051004 & 2017 May 25 & 57898.5 &  69.5 \\
\nustar\  & 3--79   & 80202051006 & 2017 Aug 30 & 57995.1 & 124.9 \\
\chandra\ & 0.3--10 & 19216       & 2017 Sep 03 & 57999.4 &  25.0 \\
\enddata                              
\end{deluxetable*}

We commenced observations of \src\ with MeerKAT on 2017 April 27.
In a typical session, following successful array configuration we
observed the calibrator PKS~1934$-$638, in order to derive and apply
the phase-delay corrections.

Since the phase stability of the array is still being investigated,
on each day we typically did three 20 minute magnetar observations
interspersed with observations of PKS~1934$-$638. From April 27 to
October 3 (a span of 159 days) we obtained a total of 231 such
observations on 74 separate days. On average these used 14.4 antennas
(on a variety of baselines dependent on availability).  This
corresponds to a cold-sky tied-array $\mbox{SEFD} \approx 32$\,Jy,
which is comparable to the equivalent Parkes telescope SEFD at L
band \citep[e.g.,][]{mlc+01}.

We preceded the magnetar observations on each day by a five minute
track on PSR~J1644$-$4559 ($P=0.45$\,s, dispersion measure
$\mbox{DM}=478$\,cm$^{-3}$\,pc), a bright pulsar with a known timing
solution, to serve as a timing calibrator.

\subsection{\xmm\ } \label{sec:xmmobs}

The \xmm\ X-ray telescope \citep{jla+01} observed \src\ on 2017
March 19 for a total of 125\,ks.  During the observation the EPIC
(European Photon Imaging Camera) pn camera \citep{sbd+01} was
operated in Full Frame mode, while the EPIC MOS cameras \citep{taa+01}
were operated in Small Window mode.  The EPIC detectors are sensitive
to energies of 0.15--15\,keV.

Standard \xmm\ data reduction threads were
used\footnote{\url{https://www.cosmos.esa.int/web/xmm-newton/sas-threads/}}
to process the data using the Science Analysis Software (SAS) version
16.  After removing the effects of soft proton flares, the usable
live time was 102\,ks.  We used a circular source extraction region
with radius 18\arcsec\ centered on the pulsar, and the background
was estimated from a circular source-free region of radius 72\arcsec.

\subsection{\swift\ } \label{sec:swiftobs}

Three observations of \src\ were made with the \swift\ X-ray Telescope
\citep[XRT;][]{bhn+05} between 2017 April 27 and May 5.  The first
observation was made in Photon Counting (PC) mode, while the others
were made in Windowed Timing (WT) mode.  PC mode gives two-dimensional
imaging capabilities at 2.5\,s resolution, while WT provides only
one spatial dimension but at a higher time resolution of 1.7\,ms.
More observational details are given in Table~\ref{tab:xrayobs}.

The observations were analyzed by running the standard XRT data
reduction pipeline \texttt{xrtpipeline} on the pulsar position (see
Table~\ref{tab:xrayeph}).  For the PC mode observation, the source
region was a circle of radius 20 pixels (0\farcm78) centered on the
pulsar, while the background region was an annulus with an inner
radius of 40 pixels and an outer radius of 60 pixels centered on
the pulsar.  For WT mode observations, the source regions were
40-pixel strips centered on the pulsar, and the background regions
were 40-pixel strips placed away from the pulsar.

\subsection{\chandra\ } \label{sec:chandraobs}

The {\em Chandra X-ray Observatory} \citep{wtvo00b} observed \src\
on 2017 May 8 and 23, and September 3, using the ACIS-S \citep{gbf+03}
spectrometer (see Table~\ref{tab:xrayobs}).  All observations were
made in Continuous Clocking (CC) mode. CC mode foregoes two-dimensional
imaging to provide 2.85\,ms time resolution.

The Chandra Interactive Analysis of Observations software
\citep[CIAO;][]{fma+06} was used to reduce the data. The data,
downloaded from the Chandra Data Archive
(CHASER\footnote{\url{http://cda.harvard.edu/chaser/}}), were first
reprocessed using the script \texttt{chandra\_repro}, and then the
appropriate science thread was
followed\footnote{\url{http://cxc.harvard.edu/ciao/threads/pointlike/}}.
Spectra were extracted using an 8-pixel (8\arcsec) strip centered
on the pulsar. The background region was a 32-pixel strip placed
away from the pulsar. Photon arrival times were corrected to the
solar system barycenter using the pulsar position.

\subsection{\nustar\ } \label{sec:nustarobs}

\src\ was observed with \nustar{} \citep[Nuclear Spectroscopic
Telescope Array;][]{hcc+13} on 2017 May 7 and 25, and August 30
(see Table~\ref{tab:xrayobs}).  These observations were coordinated
with \chandra\ (see Section~\ref{sec:chandraobs}), in order to probe
the pulsar over a broad energy range.

The data were processed using the standard HEASOFT tools
\texttt{nupipeline} and \texttt{nuproducts}, following the \nustar{}
Quickstart
Guide\footnote{\url{https://heasarc.gsfc.nasa.gov/docs/nustar/analysis/nustar_quickstart_guide.pdf}}.
Spectra from both Focal Plane Modules (FPMA and FPMB) were fit
jointly during the analysis.  Source regions were chosen to be
circles with radii of 20 pixels (8\arcmin) centered on the pulsar.
Background regions were circles of the same radius, but placed away
from the pulsar.  Event arrival times were corrected to the solar
system barycenter using the pulsar position.

\section{Analysis and Results} \label{sec:res}

\subsection{Radio Reactivation} \label{sec:reactivation}

\begin{figure}[!tp]
\centering
\includegraphics[width=0.92\linewidth]{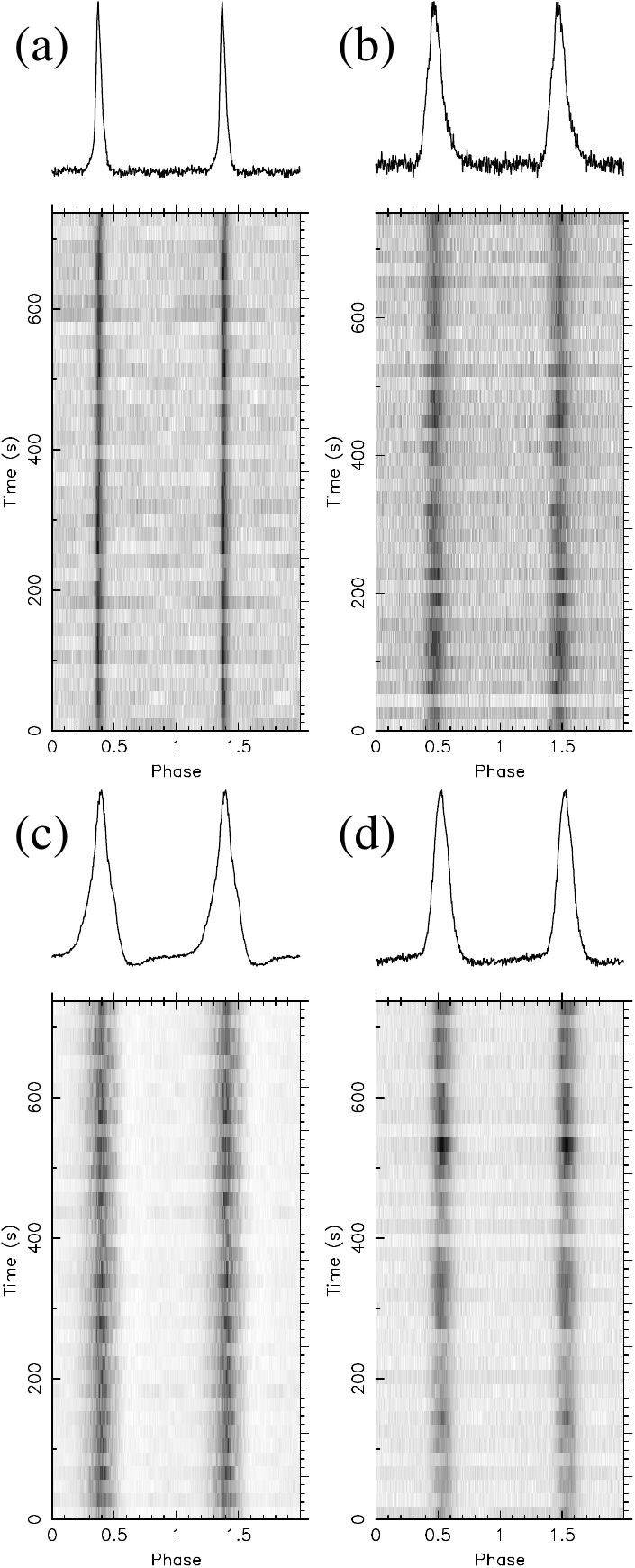}
\caption{Radio pulse profiles of \src. (a) Parkes at 3.1\,GHz
(bandwidth $\mbox{BW} = 1$\,GHz) on 2017 April 5.  (b) MeerKAT at
1.3\,GHz (RFI-free $\mbox{BW} \approx 500$\,MHz) on April 27. (c)
Parkes at 3.1\,GHz ($\mbox{BW} = 1$\,GHz) on July 4. (d) Parkes at
1.4\,GHz ($\mbox{BW}=256$\,MHz) on June 7. All profiles are displayed
(in arbitrary units) as a function of time and summed at the top,
repeated twice.  The baselines of the profiles in panels (c) and
(d) are affected by instrumental artifacts.
\label{fig:4profs}}
\end{figure}

After recognizing on 2017 April 26 that \src\ was emitting radio
waves, we inspected the previously acquired Parkes monitoring data
(Section~\ref{sec:mon}).  Data collected on April 5 showed that the
pulsar had turned on (see Figure~\ref{fig:4profs}a).  Our only other
prior unpublished Parkes monitoring observations, from 2016 November
17 and 2017 January 14, show no evidence of pulsations, with a
5\,$\sigma$ flux density limit of $\approx 0.3$\,mJy at 3\,GHz
\citep[see][]{scs+17}.

\subsection{Polarimetry} \label{sec:polresults}

\begin{figure}[!tp]
\centering
\includegraphics[width=\linewidth]{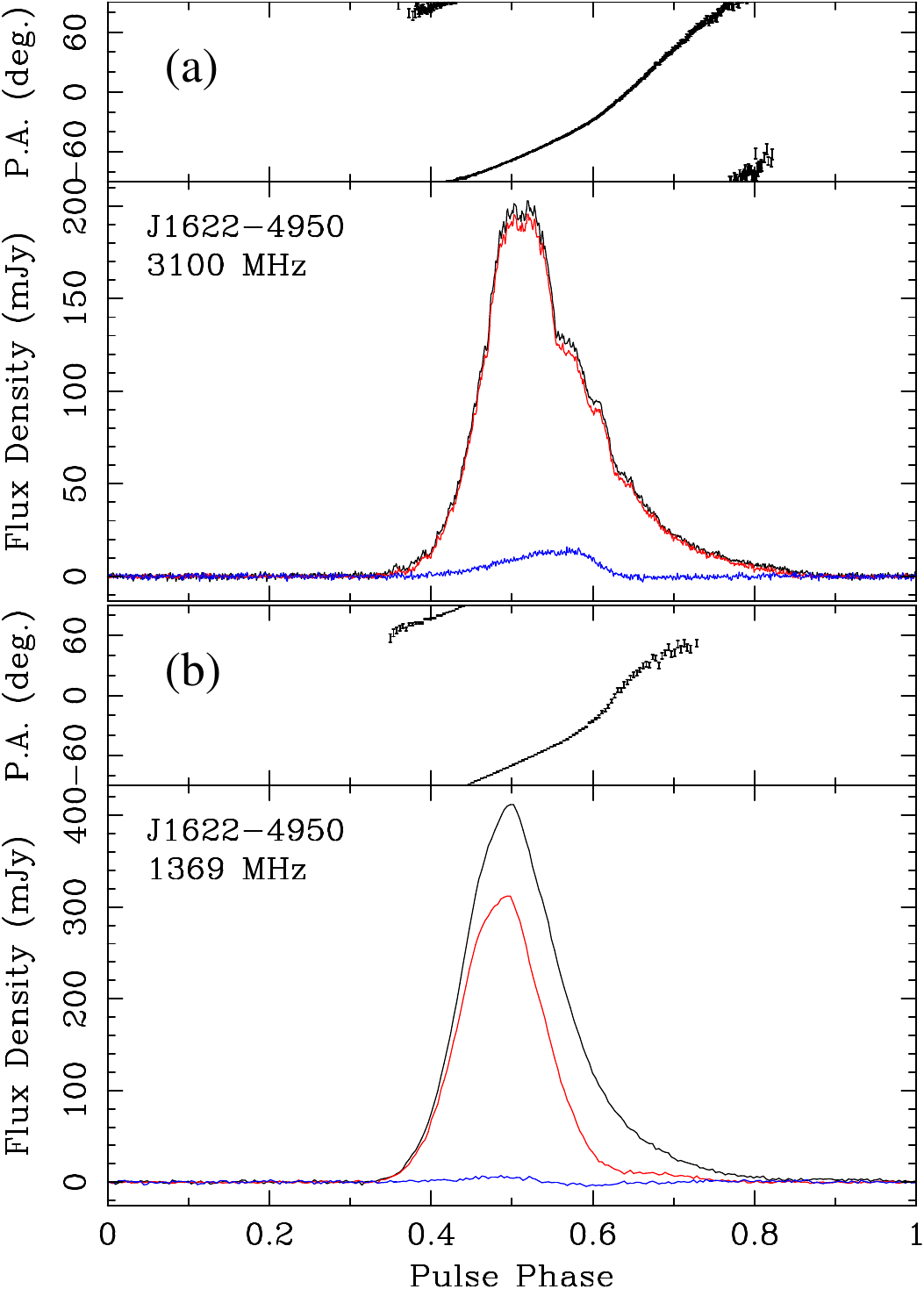}
\caption{Calibrated full-Stokes profiles of \src\ based on Parkes
observations.  The red and blue lines represent the amount of linear
and circular polarization, respectively, and the black traces show
total intensity.  The position angles (PAs) of linear polarization
have been corrected for rotation measure
$\mbox{RM}=-1495\pm5$\,rad\,m$^{-2}$ and are plotted de-rotated to
infinite frequency.  The profiles, from 2017 August 5 (panel (b))
and August 16 (panel (a)), are aligned by eye.
\label{fig:pol}}
\end{figure}

The fold-mode data on \src\ collected with PDFB4 in 2017
(Section~\ref{sec:pol}) were analyzed with PSRCHIVE \citep{hvm04}
as in, e.g., \citet{crj+07}.  The individually determined rotation
measures (RMs) are essentially consistent with the value published
in \citet{lbb+10}, and the calibrated pulse profiles are shown in
Figure~\ref{fig:pol}.  The period-averaged flux densities for these
two observations (which are the only flux-calibrated radio observations
presented in this paper) are $S_{1.4}=63\pm6$\,mJy and $S_{3}=32\pm3$\,mJy,
respectively.

\subsection{Radio Timing} \label{sec:radioTiming}

The MeerKAT observations (Section~\ref{sec:mkpsrobs}) were used to
obtain timing solutions for two purposes: to fold the \chandra\ and
\nustar\ X-ray photons, and to probe the short-term variability of
the spin-down torque on the neutron star.

All MeerKAT observations were processed in a homogeneous way using
PSRCHIVE. The first few days of Parkes and MeerKAT observations
(see Figure~\ref{fig:4profs}b) were used to improve the rotational
ephemeris thereafter used to fold all MeerKAT observations.

Radio frequency interference (RFI) was excised in a multi-step
process: first, 10\% of the recorded band was removed due to bandpass
roll-off, yielding the nominal 900--1670\,MHz MeerKAT L band; next,
a mask with known persistent RFI signals (e.g., GSM, GPS) was applied
to the data; finally, PSRCHIVE tools and clean.py, an RFI-excision
script provided by the CoastGuard data analysis pipeline \citep{lkg+16},
were used to remove remaining RFI.  More than 400\,MHz of clean
band was typically retained after this flagging.

We summed all frequency channels, integrations, and polarizations
for each observation to obtain total-intensity profiles.  We then
obtained times-of-arrival (TOAs) for each observation (i.e., two
or three per day) by cross-correlating individual profiles with a
standard template based on a very high signal-to-noise ratio
observation.

The same procedure was applied to MeerKAT observations of
PSR~J1644$-$4559. Using the TEMPO software we confirmed that the
timing solution derived for PSR~J1644$-$4559 was consistent with
published parameters \citep[see][]{mhth05}.

Like many magnetars, \src\ displays unsteady rotation
\citep[e.g.,][]{dk14}. Nevertheless, we were able to obtain a simple
phase-connected timing solution for the three week period spanning
the first two sets of paired \chandra\ and \nustar\ observations
(Table~\ref{tab:xrayobs}), fitting the TOAs with TEMPO to a model
containing only pulse phase, rotation frequency ($\nu$), and frequency
derivative ($\dot \nu$). This solution is presented in
Table~\ref{tab:xrayeph}.

\begin{deluxetable}{lc}[!tp]
\tablecaption{\src\ Ephemeris Used to Fold the X-Ray Data  \label{tab:xrayeph}}
\tablewidth{0pt}
\tabletypesize{\footnotesize}
\tablehead{
\colhead{Parameter} & \colhead{Value}
}
\startdata
R.A. (J2000)\tablenotemark{a} & $16^{\mathrm h}22^{\mathrm m}44\fs89$ \\
Decl. (J2000)\tablenotemark{a} & $-49\arcdeg50\arcmin52\farcs7$ \\
Spin frequency, $\nu\ (\text{s}^{-1})$ & $0.231090389(2)$ \\
Frequency derivative, $\dot{\nu}\ (\text{s}^{-2})$ & $-7.94(2)\times 10^{-13}$ \\
Epoch of frequency (MJD TDB) & 57881 \\
Data span (MJD) & 57880--57903 \\
Number of TOAs & 59 \\
rms residual (phase) & 0.005 \\
\hline
\multicolumn{2}{c}{Derived Parameters} \\
\hline
Spin-down luminosity, $\dot E$ (erg\,s$^{-1}$) & $7.2\times10^{33}$ \\  
Surface dipolar magnetic field, $B$ (G) & $2.6\times10^{14}$ \\
Characteristic age, $\tau_c$ (kyr) & 4.6 \\
\enddata
\tablecomments{Numbers in parentheses are TEMPO $1\,\sigma$ uncertainties. }
\tablenotetext{a}{Values fixed to those from \citet{ags+12}. }
\end{deluxetable}

\begin{figure}[!tp]
\centering
\includegraphics[width=\linewidth]{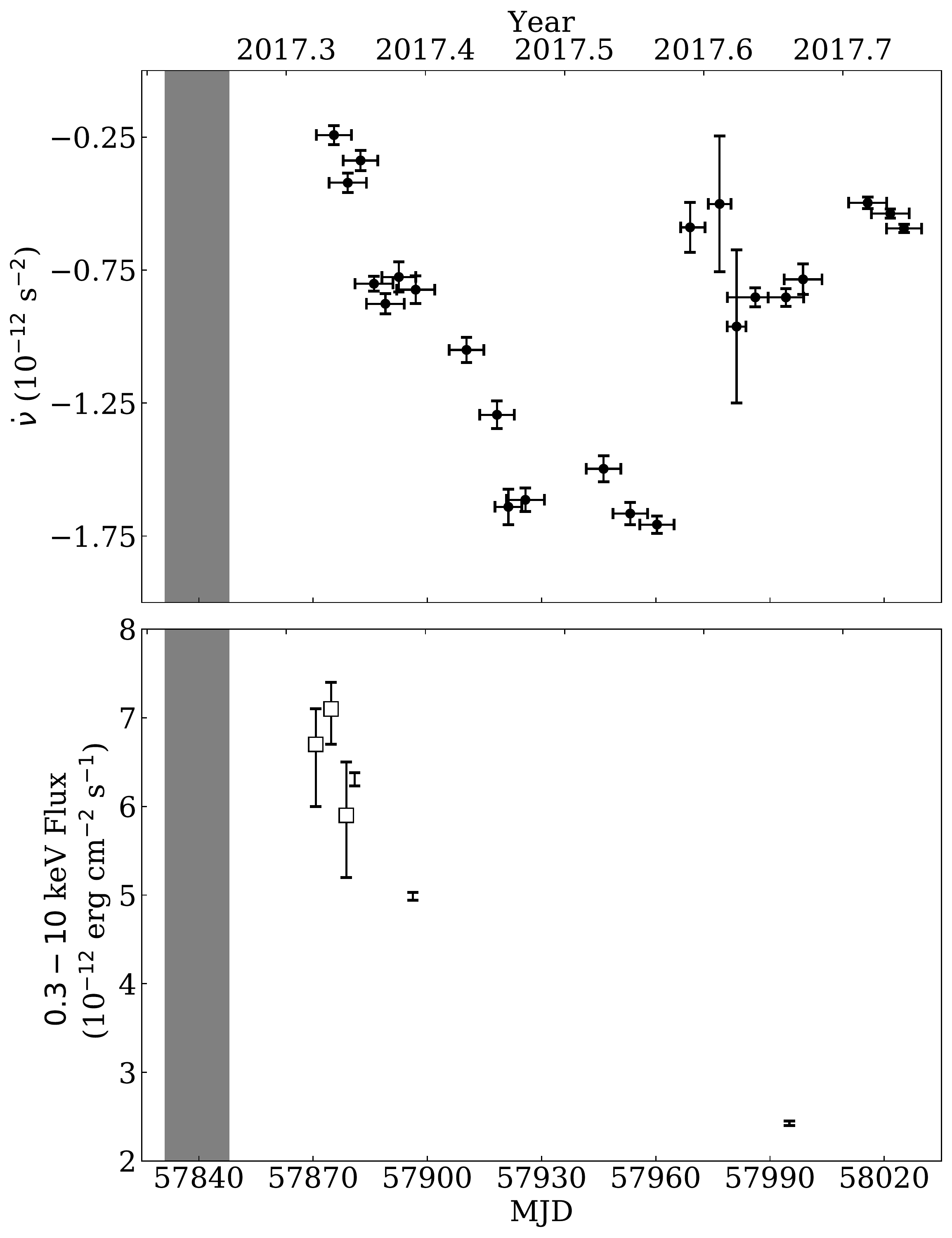}
\caption{Time evolution of properties of \src. Top panel: frequency
derivative measured from short-term timing solutions (spanning
horizontal error bars) obtained from MeerKAT data (see
Section~\ref{sec:radioTiming}).  Bottom panel: 0.3--10\,keV absorbed
X-ray flux as measured from the best-fit blackbody model (see
Table~\ref{tab:xrayfits}).  \swift\ observations are shown as open
squares and combined \chandra\ and \nustar\ observations are
represented simply by their error bars. The X-ray flux appears to
be decaying exponentially with a time constant of $111\pm19$ days
(see Section~\ref{sec:xraydisc}). All error bars represent $1\,\sigma$
confidence levels.  The gray band in both panels represents the
timespan during which the X-ray outburst and radio revival most
likely occurred (see Section~\ref{sec:disc}).
\label{fig:nudot}}
\end{figure}

Over the 5-month span of all the radio timing observations presented
here, the $\dot \nu$ for \src\ has changed by a factor of 7, and
$\ddot \nu$ has changed sign.  In order to probe the evolution of
the spin-down torque ($\propto \dot \nu$), we fit short-term
overlapping timing models where a fit for $\nu$ and $\dot\nu$ proves
adequate, i.e., with featureless timing residuals.  Each of these
short-term timing solutions spanned approximately one week.
Figure~\ref{fig:nudot} shows the run of $\dot \nu$ from these
solutions spanning five months.

\subsection{Pre-outburst \xmm\ Limit} \label{sec:xmmlimit}

In the 2017 March 19 \xmm\ observation, performed before the April
5 observation that detected resumed radio activity from \src, we
detect no X-ray counts in excess of the background rate.  Using the
Bayesian method of \citet{kbn91}, we place an upper limit of
0.002\,s$^{-1}$ on the background-subtracted EPIC-pn 0.3--10\,keV
count-rate at a $5\,\sigma$ confidence level. (We use only the pn
data to place an upper limit as the MOS detectors are much less
sensitive.) For this limit, {\tt WebPIMMS} gives a 0.3--10\,keV
absorbed flux limit of $9\times10^{-15}$\,erg\,cm$^{-2}$\,s$^{-1}$,
assuming an absorbed blackbody (BB) spectrum with $kT=0.4$\,keV
\citep[typical of a quiescent magnetar;][]{ozv+13} and $N_\text{H}
= 6.4\times 10^{22}\, \text{cm}^{-2}$ (see Section~\ref{sec:xrayfits}).
The corresponding $5\,\sigma$ unabsorbed flux limit is
$8\times10^{-14}$\,erg\,cm$^{-2}$\,s$^{-1}$.

\subsection{X-Ray Pulsations} \label{sec:xraypulse}

\begin{figure}[!tp]
\centering
\includegraphics[width=\linewidth]{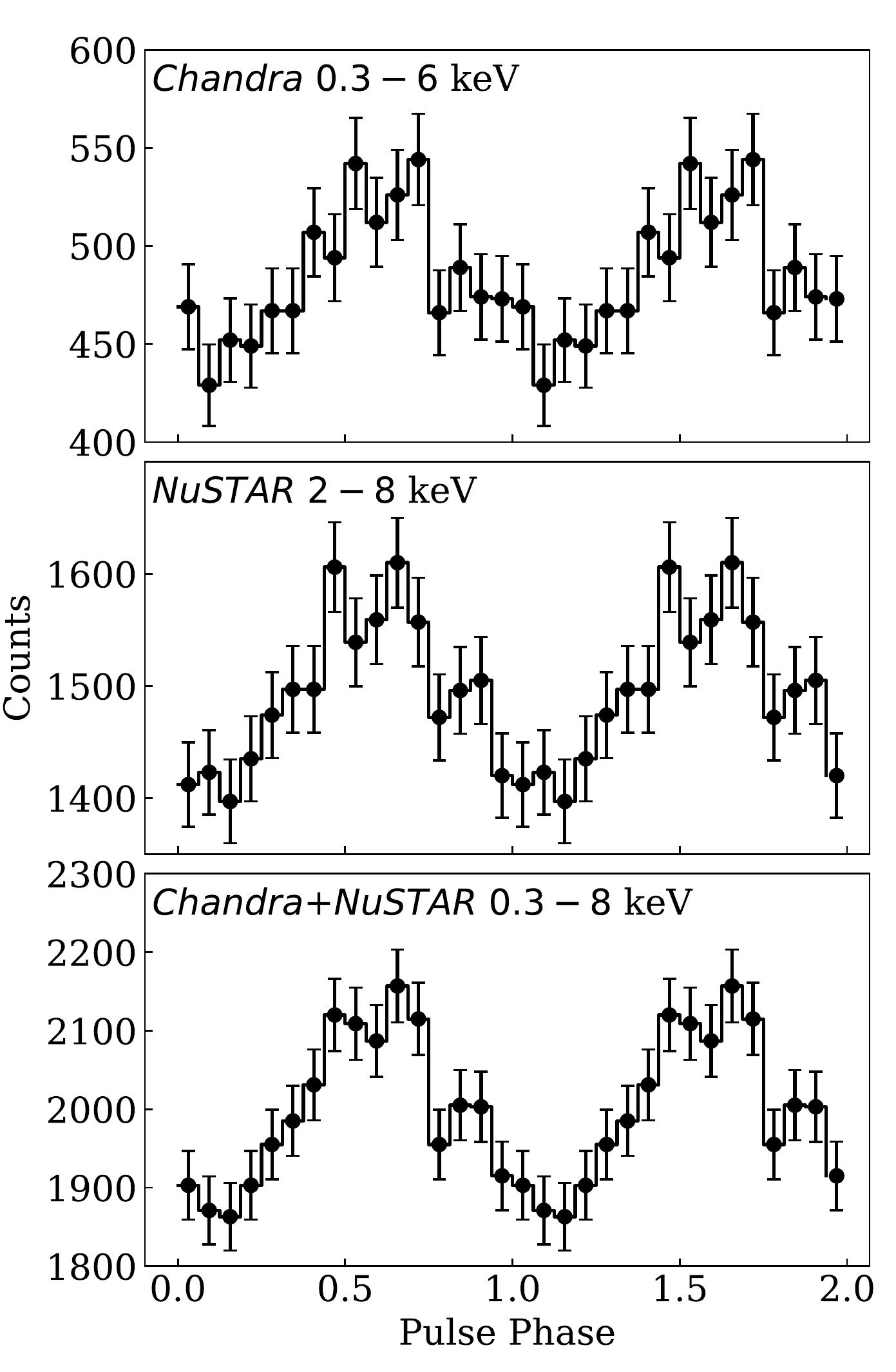}
\caption{X-ray pulse profiles of \src. Data from the first two sets
of paired \chandra\ and \nustar\ observations were folded with the
ephemeris from Table~\ref{tab:xrayeph} (see Section~\ref{sec:xraypulse}).
The top two panels show individual instrument profiles in the noted
energy ranges. The bottom panel shows the overall combined profile.
\label{fig:folds}}
\end{figure}

We have used the radio-derived rotational ephemeris presented in
Table~\ref{tab:xrayeph} to fold barycentered photons from the first
two sets of paired \chandra\ and \nustar\ observations, and we
clearly detect pulsations (Figure~\ref{fig:folds}).  To maximize
the significance of the detected pulsations, we optimize the energy
range that maximizes the $H$ statistic of the pulse \citep{dsr89}.
For \chandra\ we find the optimal range to be 0.3--6.0\,keV, with
a false alarm probability, given the maximum value of the $H$
statistic, of $P_{\rm fa}=2\times10^{-5}$ (equivalent to a $4.3\,\sigma$
detection).  For \nustar\ we find the optimal range to be 2.0--8.0\,keV
($P_{\rm fa}=1\times10^{-7}$; $5.3\,\sigma$).  We also present a
combined \chandra\ and \nustar\ 0.3--8\,keV profile which shows a
strong pulse ($P_{\rm fa}=5\times10^{-12}$; $6.9\,\sigma$).  Note,
however, that \nustar\ is not sensitive to X-ray photons below
$\sim2$\,keV.  This is the first detection of X-ray pulses from
\src. The pulsed fraction (PF) of the 0.3--6\,keV \chandra\ profile
is $\mbox{PF} =5\%\pm1\%$ (using the RMS method described in
\citealt{aah+15} Appendix A, where PF errors are reported at the
$1\,\sigma$ level).  For the 2--8\,keV \nustar\ pulsations, $\mbox{PF}
= 4.1\%\pm0.7\%$.  This is far below the $3\,\sigma$ upper limit
of $\mbox{PF} < 70\%$ at 0.3--4\,keV determined by \citet{ags+12}.

\begin{deluxetable*}{ccccc}[!t]
\tablecaption{Spectral Fits for X-Ray Detections of \src\ in 2017 \label{tab:xrayfits}}
\tablecolumns{5}
\tablewidth{0pt}
\tabletypesize{\footnotesize}
\tablehead{
	\colhead{Telescope} & \colhead{ObsID} & \colhead{Start Time} & \colhead{0.3--10\,keV Absorbed Flux} & \colhead{$kT$} \\
	\colhead{} & \colhead{} & \colhead{(MJD)} & \colhead{($10^{-12}$\,erg\,cm$^{-2}$\,s$^{-1}$)} & \colhead{(keV)}
}   
\startdata
\swift\           & 00010071001       & 57870.7 & $6.7\substack{+0.4 \\ -0.7}$    & $0.80\pm0.04$ \\
\swift\           & 00010071002       & 57874.7 & $7.1\substack{+0.3 \\ -0.4}$    & $0.77\pm0.03$ \\
\swift\           & 00010071003       & 57878.7 & $5.9\substack{+0.6 \\ -0.7}$    & $0.79\substack{+0.07 \\ -0.06}$ \\
\nustar/\chandra\ & 80202051002/19214 & 57880.9 & $6.30\substack{+0.08 \\ -0.07}$ & $0.791\pm0.006$ \\
\nustar/\chandra\ & 80202051004/19215 & 57896.2 & $4.98\substack{+0.05 \\ -0.04}$ & $0.770\pm0.005$ \\
\nustar/\chandra\ & 80202051006/19216 & 57995.1 & $2.42\substack{+0.03 \\ -0.02}$ & $0.778\pm0.006$ \\
\enddata
\tablecomments{This joint absorbed blackbody fit (\texttt{tbabs*bbody})
yielded a Cash statistic of 6201 and $\chi^2 = 5932$ for 6115 degrees
of freedom (reduced $\chi^2 = 0.97$). All uncertainties are given
at the $1\,\sigma$ confidence level. The absorbing column density,
$N_\mathrm{H} = (6.4\pm0.1) \times 10^{22}$\,cm$^{-2}$, was constrained
to have the same value for every observing epoch. See
Section~\ref{sec:xrayfits} for more details.
}
\end{deluxetable*}

To probe for PF variability, we also measured the pulsed fraction
in each individual \chandra\ and \nustar\ observation.  As above,
for the first two \chandra\ and \nustar\ epochs we folded each
observation using the ephemeris in Table~\ref{tab:xrayeph}: each
individual-epoch, single-telescope pulsed detection is significant
at the $\approx 3\,\sigma$ level.  This ephemeris is not valid for
the third \chandra\ and \nustar\ epochs, nor could we obtain one
phase-connected timing solution that spans all three epochs.  We
folded those photons with the ephemeris corresponding to the relevant
frequency derivative measurement presented in Figure~\ref{fig:nudot}
($\nu=0.23107824(1)$\,s$^{-1}$, $\dot\nu=-7.8(5)\times10^{-13}$\,s$^{-2}$
at an epoch of MJD~57994).  The third \nustar\ observation yielded
a $3.1\,\sigma$ pulsed detection, while the third \chandra\ observation
did not result in a significant detection.  The corresponding PFs,
in chronological order, are $3\%\pm1\%$, $4\%\pm1\%$, and $4\%\pm1\%$
at 2--8\,keV for the \nustar\ observations, and $4\%\pm2\%$,
$5\%\pm2\%$, and $<10\%$ ($3\,\sigma$) at 0.3--6\,keV for the
\chandra\ observations.  Thus, we do not find significant PF evolution
within the 120 days probed by our measurements.

\subsection{Spectral Analysis} \label{sec:xrayfits}

The \swift, \chandra, and \nustar\ spectra were fit with
\texttt{XSPEC}\footnote{\label{fn:xspec}\url{https://heasarc.gsfc.nasa.gov/xanadu/xspec/}}.
We fit several photoelectrically absorbed models using Tuebingen-Boulder
absorption \citep[{\tt tbabs} in {\tt XSPEC};][]{wam00} and
\citet{cash79} statistics. The abundances from \citet{wam00} and
the photoelectric cross sections from \citet{vfky96} were used.  We
treated each pair of closely spaced \chandra\ and \nustar\ observations
as a single dataset in the spectral fitting.  We first fit the
\swift\ and \chandra+\nustar\ spectra with individual BB and power-law
(PL) models. The low count rate prohibited resolving separate model
components in the \swift\ observations, and we fit a BB+PL model
to only the \chandra+\nustar{} datasets.  In each of the fits, the
absorbing hydrogen column density $N_\mathrm{H}$ was constrained
to have the same value across all epochs, but the BB temperature
$kT$ and photon index $\Gamma$ were allowed to vary from epoch to
epoch. All three models provided acceptable fits, with a reduced
$\chi^2$ value of 0.97 for the BB model, 0.99 for PL, and 0.92 for
BB+PL.  Since the single-component models produced acceptable fits,
which were not significantly improved by the addition of the extra
component, we do not present the BB+PL model.  Additionally, the
PL model yields $\Gamma = 5$ for all observations (with uncertainties
ranging from 0.04 to 0.4), which is large for a magnetar.  The
$N_\mathrm{H} = (1.66\pm0.02) \times10^{23}$\,cm$^{-2}$ for that
model is large, and is likely the result of the fitted absorption
compensating for the model's soft X-ray flux that is not intrinsic
to the source.  This $N_\mathrm{H}$ would also be an outlier to the
empirical DM--$N_\mathrm{H}$ relation of \citet{hnk13}.  By contrast,
the BB model yields $N_\mathrm{H} = (6.4\pm0.1) \times 10^{22}$\,cm$^{-2}$,
which fits this relationship, and the $kT$ values (see
Table~\ref{tab:xrayfits}) are much more typical of a magnetar in
outburst.  We therefore conclude that the BB model provides the
best description of the spectra.  The results of the BB spectral
fits are summarized in Table~\ref{tab:xrayfits}, and Figure~\ref{fig:nudot}
shows the time evolution of the absorbed X-ray flux.

As is evident from our spectral fits, which require no additional
PL component, we detect very little emission above 10\,keV.  We
probed for a hard PL component by searching for an excess of counts
above 15\,keV in the first \nustar\ observation when the source was
brightest. We find that the number of counts is consistent with the
background and place a $5\,\sigma$ limit on the 15--60\,keV count
rate of 0.002\,s$^{-1}$ \citep[we choose this energy range to be
consistent with the literature; e.g.,][]{esk+17}. This implies an
absorbed 15--60\,keV flux limit of
$9\times10^{-13}$\,erg\,cm$^{-2}$\,s$^{-1}$, using the measured
$N_\text{H} = 6.4\times 10^{22}\,\text{cm}^{-2}$ and assuming a PL
index of $\Gamma=1$, typical for the hard X-ray component of magnetars
\citep[e.g.,][]{ahk+13,vhk+14}.

A trend of decreasing X-ray flux is evident from the observations
made thus far (see Figure~\ref{fig:nudot}).  The peak absorbed flux
values are $\approx 800\times$ greater than the \xmm{}\ limit on
2017 March 19 (Section~\ref{sec:xmmlimit}).  This enormous increase
in flux shows that \src\ has recently gone into a phase of X-ray
outburst.  So far, the BB temperature $kT$ is consistent with being
constant.

\section{Discussion} \label{sec:disc}

After being dormant for 2--3 years, \src\ resumed radio emission
between 2017 January 14 and April 5 (Section~\ref{sec:reactivation}).
In turn, its X-ray flux increased by a factor of at least 800 between
2017 March 19 and April 27 (Section~\ref{sec:xmmlimit} and
Table~\ref{tab:xrayfits}).  Transient radio emission from magnetars
has been shown to be associated with X-ray outbursts
\citep[e.g.,][]{crhr07,hgr+08}.  The most recent outburst of \src\
therefore most likely happened between 2017 March 19 and April 5.
This provides the opportunity to study the behavior of this magnetar
soon after outburst, and compare it to that previously observed as
well as with that of other magnetars.

\subsection{Radio Variability and Outburst History}

The previous X-ray outburst of \src\ is thought to have occurred
in the first half of 2007 \citep{ags+12}, and radio emission following
that outburst \citep[retrospectively detected in 2008, following
the discovery of the magnetar in 2009;][]{ags+12} became undetectable
in 2015 \citep{scs+17}. The outburst history prior to 2007 is not
as well constrained, but on the basis of the radio behavior it is
consistent with an outburst having occurred in or before 1999, and
radio emission becoming undetectable no earlier than 2004
\citep[see][particularly their Figure 3]{scs+17}.

Following the 2007 outburst, the radio pulse profiles of \src\
displayed great variability, covering up to 60\% of pulse phase
\citep{lbb+10,lbb+12}. In 2017, the observed profiles display
variability (see Figures~\ref{fig:4profs}a and \ref{fig:4profs}c),
although not yet as great or covering such a large range of rotational
phase (Figures~\ref{fig:4profs} and \ref{fig:pol}). Also, most of
our radio profiles in 2017 are from MeerKAT at 0.9--1.7\,GHz, and
at low frequencies the noticeably scatter-broadened profiles (compare
Figures~\ref{fig:4profs}b and \ref{fig:4profs}d) mask smaller
variability.

The 3\,GHz flux density measured in 2017, 32\,mJy (Figure~\ref{fig:pol}a
and Section~\ref{sec:polresults}), is $100\times$ larger than the
$S_{3} \approx 0.3$\,mJy limits during 2015--2016 \citep{scs+17}
and early 2017 (Section~\ref{sec:reactivation}).  While the flux
densities for radio magnetars are known to fluctuate greatly (due
to a combination of changing pulse profiles and varying flux density
from particular profile components), in the two years prior to
disappearance by 2015, $S_{3}$ for \src\ decreased on a trend from
$\approx 10$\,mJy to $\sim 1$\,mJy \citep{scs+17}, and the one
current flux-calibrated value of $S_{3}$ is larger by a factor of
about 2 than any reported before for this magnetar.  Likewise, the
one current flux-calibrated measurement at 1.4\,GHz, $S_{1.4}=63$\,mJy
(Figure~\ref{fig:pol}b and Section~\ref{sec:polresults}), is the
largest such value ever reported for this magnetar \citep[and
comparable to the maximum non-calibrated values from 2000 to 2001;
see Figure 3 of][]{scs+17}.  Thus, while the flux densities are
currently fluctuating, these initial measurements together with the
historical record are compatible with the notion that relatively
soon after outburst, \src\ reaches maximum period-averaged flux
densities at 1.4--3\,GHz of tens of mJy --- and apparently not
substantially more or less (with the caveat that we started observing
within one month of the latest outburst, while radio observations
started only two years after the 2007 outburst).

Unlike ordinary pulsars, the radio spectra of magnetars are remarkably
flat, resulting in pulsed detections at record frequencies of nearly
300\,GHz \citep[e.g.,][]{tde+17}.  Owing to the varying flux
densities, the reliable determination of magnetar spectra ideally
requires simultaneous multi-frequency observations.  \citet{pmp+17}
report on 2.3 and 8.4\,GHz ``single polarization mode'' observations
of \src\ with the Deep Space Network DSS-43 antenna on 2017 May 23,
from which they obtain a spectral index $\alpha_{2.3-8.4}=-1.7\pm0.2$
(where $S_f \propto f^{\alpha}$).  Without further details we cannot
assess whether the single polarization mode might bias the flux
density determination in a highly polarized source.  In any case,
their reported mean $S_{2.3} = 3.8\pm0.8$\,mJy is a factor of $\sim
10$ smaller than our two Parkes flux-calibrated measurements. Our
own flux-calibrated measurements presented in this paper correspond
to a nominal spectral index $\alpha_{1.4-3.1} = -0.8$, but these
were not performed simultaneously. Either of these $\alpha$ values
would correspond to steeper spectra than have been measured for
this pulsar \citep[e.g.,][]{ags+12}, and further investigations are
desirable.

\subsection{Polarimetry and Magnetospheric Geometry}

\citet{lbb+12} showed that the \src\ radio profile changes significantly
from observation to observation but they were able to classify the
various profiles into four main types (see their Figure 3). The
present profile at 1.4\,GHz (Figure~\ref{fig:pol}b) does not resemble
any of these categories.  In particular the trailing component
appears to now be completely suppressed, the circular polarization
is opposite in sign, and the linear polarization fraction is now
somewhat larger (although still much less than the 3\,GHz fraction,
in part presumably due to interstellar scattering; see
\citealt{crj+08,lbb+12}).  The RVM fits to the data in
Figure~\ref{fig:pol}a give $\alpha \approx 20\degr$ (angle between
the magnetic and rotation axes) and $\beta \approx -10\degr$ (angle
of closest approach of the line of sight to the magnetic axis).
This is broadly the same geometry as in \citet{lbb+12}.  However,
the location of the inflection point in position angle (PA) has
changed substantially.  Whereas in \citet{lbb+12} the inflection
point occurred prior to the observed profile peak (see, e.g., their
Figure~4), it now comes substantially later than the peak (near
phase 0.65 in Figure~\ref{fig:pol}a); the difference between the
inflection points is $92\degr\pm5\degr$.  This can also be seen by
the fact that the observed PA swing now appears largely concave
whereas previously it was convex. Our conclusion is that we are now
seeing emission from a very different location in the magnetosphere
compared to previously.

\subsection{X-Ray Pulsations}

We have presented the first detection of X-ray pulsations from \src\
(Figure~\ref{fig:folds}).  The profile appears to be a broad sinusoid,
with a small amount of higher harmonic structure seen as a secondary
peak on its trailing edge. Such low harmonic content is common in
magnetars \citep{kb17}.

The measured pulsed fraction for \src, $\mbox{PF} = 5\%$, appears
to be low for a magnetar not in quiescence (more typical values are
$\approx 30$\%; \citealt{kb17}), and does not yet appear to be
changing as the magnetar cools following its recent outburst.
However, such low PFs both following outbursts and in quiescence
have been observed in some magnetars. \citet{sk11} measured a PF
as low as $6\%\pm2\%$ immediately following the 2009 outburst of
1E~1547.0$-$5408, which then increased as the magnetar's flux
decreased. In quiescence, the magnetar 4U~0142+61 has $\mbox{PF} =
5\%$. Interestingly, following both its 2011 and 2015 outbursts,
the PF increased following the outburst and decreased back to the
quiescent value on a timescale of approximately one month \citep{aks+17}.
Since we did not observe the first month of the recent outburst of
\src, we cannot determine whether a post-outburst increase or
decrease occurred, nor  whether the low PF that we have measured
is similar to its quiescent value.  However, further sensitive
observations in the coming months could constrain the PF evolution
for \src.

\subsection{X-Ray Flux and Spectral Evolution} \label{sec:xraydisc} 

With {\tt XSPEC}, we infer an unabsorbed 0.3--10\,keV flux for the
first \chandra\ detection in 2017 (cf.\ Table~\ref{tab:xrayfits})
of ($1.6\pm0.1)\times10^{-11}$\,erg\,cm$^{-2}$\,s$^{-1}$.  Using
the only available estimate of the distance to \src\ \citep[9\,kpc
from its DM;][]{lbb+10}, the corresponding X-ray luminosity is $L_X
\approx 1.5\times10^{35}$\,erg\,s$^{-1}$ (as usual the DM-derived
distance has a substantial but unknown uncertainty).  This far
exceeds the contemporaneous spin-down luminosity (Table~\ref{tab:xrayeph}),
i.e., $L_X \gg \dot E$. By contrast, during quiescence the unabsorbed
X-ray luminosity (Section~\ref{sec:xmmlimit}) is $L_X <
7.7\times10^{32}$\,erg\,s$^{-1}$.  The last measured value of
frequency derivative for \src\ before quiescence
\citep[$\dot\nu=-1.3\times10^{-13}$;][]{scs+17} corresponds to $\dot
E = 1.1\times10^{33}$\,erg\,s$^{-1}$.  Thus, during quiescence $L_X
\la \dot E$. These properties are characteristic of transient
magnetars.

Following an X-ray flux increase of three orders of magnitude over
its quiescent value (Section~\ref{sec:xrayfits}), the flux of \src\
is clearly waning (Figure~\ref{fig:nudot}). The flux evolution and
spectral properties of the magnetar are broadly similar to those
measured for its putative 2007 outburst.  Fitting an exponential
decay model to our measured fluxes yields an e-fold decay timescale
of $111\pm19$ days.  This is shorter than the 360 days measured by
\citet{ags+12} for the previous outburst decay, although that
timescale was measured over 1350 days starting later post outburst,
compared to 130 days now starting roughly one month after outburst.
The BB temperature $kT = 0.8$\,keV measured early during this
outburst is similar to that measured in 2007--2009 and higher than
$kT=0.5$\,keV measured in 2011 \citep{ags+12}, although their
uncertainties were larger than ours.

The BB temperature and X-ray flux decay timescale for \src\ are
similar to those measured for other magnetar outbursts.  In the
weeks to months following outbursts, transient magnetars typically
have high $kT$ \citep[$>0.7$\,keV; e.g,][]{sk11,crt+17}, compared
to their quiescent BB temperatures \citep[$kT \approx
0.4$\,keV;][]{ozv+13}. Also, post-outburst magnetar light-curves
typically decay on timescales of hundreds of days
\citep[e.g.,][]{skc14,crt+17}.  The lack of spectral evolution in
the relaxation thus far is also not particularly unusual \citep[see,
e.g.,][]{rep+13}; however, as the flux decays by more than an order
of magnitude from its maximum we expect a decline in $kT$ as it
returns to the quiescent value.

Some magnetars show spectral turnovers above $\sim 10$\,keV
\citep[e.g.,][]{khm04}, such that most of their energy output emerges
above the traditionally studied soft X-ray band.  For \src, we have
not detected any emission above 15\,keV (Section~\ref{sec:xrayfits}).
Based on the unabsorbed soft X-ray flux at the epoch of the first
joint \chandra\ and \nustar\ observations \citep[in the 1--10\,keV
range, to be consistent with the literature;][]{esk+17}, and the
limit on the hard X-ray flux, we derive a hardness ratio of $L_H/L_S
< 0.07$.

The transient magnetars SGR~0501+4516, 1E~1547.0$-$5408, SGR~1833$-$0832
\citep[][and references therein]{esk+17}, and SGR~1935+2154
\citep{ykj+17}, all showed PL components within days of their
outbursts. Subsequent observations of SGR~0501+4516 and 1E~1547.0$-$5408
indicate that the emission became softer with time.  The non-detection
of a hard X-ray component in \src\ could therefore be due to the
one to two month delay between the outburst and the first \nustar\
observation, although we cannot exclude that $L_H/L_S$ was always
small for this magnetar.

\subsection{Torque Evolution}                                                                       
Torque increase following X-ray outbursts is common in magnetars,
and a broad trend appears to be emerging from at least a subset of
them: a monotonic increase in the torque on the neutron star followed
by a period of erratic variations and finally a monotonic decrease.
Each of these phases lasts from months to a few years.

While the torques on the transients XTE~J1810$-$197 and \src\ were
not well sampled after their outbursts in 2003 and 2007, respectively,
they both showed rapid variations, followed by a gradual monotonic
decrease over a period of a few years. Both magnetars turned off
in the radio band following these torque decreases \citep{crh+16,scs+17}.
1E~1048.1$-$5937 experienced torque increases following each of its
X-ray outbursts in 2002, 2007, and 2012, which then recovered to
the pre-outburst value on a timescale of $\sim600$ days with erratic
variations in between \citep{akn+15}.  Comparable trends have been
observed following the 2008 and 2009 outbursts of 1E~1547.0$-$5408
(\citealt{dksg12}; F.\ Camilo et al.\ 2018, in preparation).

Before its radio disappearance in 2014, the torque observed for
\src\ had smoothly decreased to half of the lowest value observed
so far in 2017 \citep[][and Figure~\ref{fig:nudot}]{scs+17}.  This
period of monotonic torque variation started in 2012, some five
years after the previous outburst.  The torque increase that we
have measured following the recent outburst (peaking at a value
60\% larger than ever before observed for this magnetar) occurred
over a period of $\approx 100$ days, and is being followed by erratic
variations (Figure~\ref{fig:nudot}).  By analogy with the 2007
outburst (for which, however, a reliable torque record started only
in 2009), we are still in the phase where erratic variations could
be expected to continue for several hundred more days.

\section{Conclusions}

We have presented new radio and X-ray observations of \src\ that
demonstrate that this magnetar most likely reactivated between 2017
March 19 and April 5.  This is the first magnetar for which radio
emission has been re-detected following a long period of inactivity.
The (variable) radio flux density is approximately $100\times$
larger than during its dormant state that lasted for more than two
years. The X-ray flux one month after reactivation was at least
$800\times$ larger than during quiescence, and has been decaying
seemingly exponentially on a $\approx 100$ day timescale, with
roughly constant spectrum thus far. This high-flux state, together
with a radio-derived rotational ephemeris, have enabled for the
first time the detection of X-ray pulsations for this magnetar,
with a small pulsed fraction of 5\%.  The pulsar's geometry inferred
from a polarization analysis of the radio emission appears to be
broadly consistent with that determined six to eight years earlier.
However, the RVM model fits, and an observed change in the inflection
point of the classic PA ``S'' swing, suggest that we are now seeing
radio emission from a different location in the magnetosphere than
previously. This indicates a novel way in which radio emission from
magnetars can differ from that of ordinary pulsars.  Further
investigation of this effect could potentially open a new window
into the large-scale behavior of plasma flows and magnetic field
geometry in magnetars.  The torque on the neutron star is currently
varying rapidly and unsteadily, as is common for magnetars following
outburst, having changed by a factor of 7 within six months of
reactivation, and we expect additional such variations for several
months to come.

\acknowledgements

The Parkes Observatory is part of the Australia Telescope, which
is funded by the Commonwealth of Australia for operation as a
National Facility managed by CSIRO.

Based on observations obtained with \xmm, an ESA science mission
with instruments and contributions directly funded by ESA Member
States and NASA.

This research has made use of data obtained from the Chandra Data
Archive, and software provided by the Chandra X-ray Center (CXC)
in the application package CIAO.

This work made use of data from the \nustar\ mission, a project led
by the California Institute of Technology, managed by the Jet
Propulsion Laboratory, and funded by the National Aeronautics and
Space Administration.

We acknowledge the use of public data from the \swift\ data archive.

This research has made use of data obtained through the High Energy
Astrophysics Science Archive Research Center Online Service, provided
by the NASA Goddard Space Flight Center.

P.S.\ holds a Covington Fellowship at DRAO.
R.F.A.\ acknowledges support from an NSERC Postdoctoral Fellowship.
M.B.\ is an ARC Laureate Fellow and receives support for the ARC Centre
of Excellence for Gravitational Wave Discovery (OzGrav).

\end{document}